\documentstyle[aps,prl,amssymb,floats,epsf,psfig]{revtex}

\newcommand{\be}{\begin{equation}}
\newcommand{\bel}[1]{\begin{equation}\label{#1}}
\newcommand{\ee}{\end{equation}}
\newcommand{\bea}{\begin{eqnarray}}
\newcommand{\ba}{\begin{array}}
\newcommand{\eea}{\end{eqnarray}}
\newcommand{\ea}{\end{array}}

\begin{document}

\twocolumn[\hsize\textwidth\columnwidth\hsize\csname@twocolumnfalse%
\endcsname

\title{Boundary induced phase transitions in driven lattice gases with
 meta-stable states}
\author{C{\'{e}}cile Appert and  Ludger Santen}

\address{ CNRS-Laboratoire de Physique Statistique, Ecole Normale
Sup{\'{e}}rieure, 24, rue Lhomond, F-75231 Paris Cedex 05, France}

\date{\today}

\maketitle

\begin{abstract}
We study the effect of meta-stability onto boundary induced phase transitions
in a driven lattice gas.
The phase diagram for open systems,
parameterized by the input and output rates, consists of two regions
corresponding to the free flow and jammed phase. Both have been
entirely characterized. The microscopic states in the high density
phase are shown to have an interesting striped structure, which
undergoes a coarsening process, and
survives in the thermodynamic limit. 
\end{abstract}
]

Driven lattice gases (DLG) are characterized by a non-vanishing mass-flow in
the stationary state which is generated by non-potential forces
\cite{sz}. 
This feature of DLG has far reaching consequences, because even the
stationary state of the system is not described in the framework of
standard equilibrium statistical mechanics \cite{gunter}. In recent years
the study of various non-equilibrium lattice gas models, which have
important applications, e.g. in biological context \cite{MacD69} or as
models for traffic flow \cite{NaSch92,CSS99}, led to a profound
understanding of some key features of systems far from equilibrium.

One of the most fascinating effects observed for this kind of models 
are so-called boundary induced phase transitions, i.e. the change of
the bulk properties due to a variation of the  boundary conditions
\cite{Krug91,Popkov00}. 
Such phase transitions have been extensively studied 
for a class of 1D DLG with a continuous
flow density relation for periodic boundary conditions (PBC).
For some models even {\em exact} results for the stationary state
exist
\cite{Gunter93,HonRaj,EvansdeGier}.
A phenomenological theory \cite{Kolo98} was
recently
able to predict the phase diagram 
in the case of open boundary conditions (OBC) also for more
complicated models \cite{Popkov99}. 
However, less is known for DLG which exhibit {\em
meta-stable states} for PBC \cite{Barlo}, although they
are relevant in different fields. Here, we study one of the simplest
DLG model which shows meta-stability for PBC.
We first discuss briefly the case of PBC, but our main focus is 
to study the effect of meta-stability in the case of OBC. Using
a combination of analytical and numerical methods we explore the 
structure of the phase diagram and characterize the typical behavior
in the different phases.

We study a DLG model defined on a chain of length $L$. 
Each site of the chain may either be empty ($x$) or
occupied with a particle of velocity zero ($0$) or one ($1$). 
In a first step all particles which moved in the previous time step
keep their velocity {\em if} the next site is empty.
With probability $q_0$, velocity
one is assigned to particles with velocity zero and an empty site ahead.
In all other cases velocity zero is
assigned to the particle. In the second step particles with velocity
one move to the next site. Both steps are applied synchronously to all
particles.   

This model may be interpreted as a special case of the recently
introduced {\em VDR} model for traffic flow \cite{Barlo}, i.e. we
restrict the maximal velocity of the particles to one. For $q_0 =
1$ we recover the deterministic asymmetric exclusion process with
parallel dynamics \cite{EvansdeGier}.

\begin{figure}[h]
\centerline{\psfig{figure=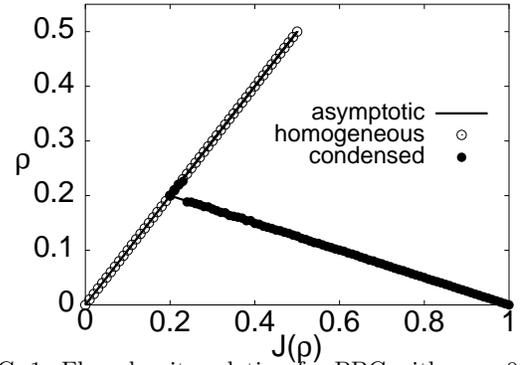,height=5cm}}
\caption{\protect{Flow density relation for PBC with 
$q_0 = 0.25$ and $L= 1,000$. Empty circles show simulation results
for an initialization where all particles have velocity one and at
least one empty site in front. Filled circles show
results  starting the simulation with a compact cluster of
particles. These results are compared to our estimate for $L\to \infty$
(solid line).}}
\label{fig_periodic}
\end{figure}

Following \cite{Kolo98}, the structure of the flow-density
relation, i.e. the fundamental diagram, in the case of PBC
is a key to understand the behavior of the open chain.
Fig.~\ref{fig_periodic} shows the numerically established 
fundamental diagram for our model, in comparison with:

\begin{eqnarray*}
 J(\rho) & = & \rho         \quad \text{ if } 0\leq \rho \leq 0.5  \\
 J(\rho) & = & q_0 (1-\rho)  \quad \text{ if } q_0/(1+q_0)\leq \rho \leq 1  
\end{eqnarray*}

The form of the fundamental diagram can be easily understood. 
For densities $0\leq \rho \leq 0.5$ and suitable initial
configurations (the initial velocity of the particles is one and
no particle is blocked) all particles move deterministically with 
velocity one. At larger densities $\rho > 0.5$ the number of particles
exceeds the number of holes, i.e. one cannot avoid the formation  of
clusters of particles. The simulation results indicate that in the
steady state one observes phase separation, i.e. the formation of a large compact
cluster of particles which coexists with a free-flow regime. 
The density in the free-flow regime is determined by the
frequency of particles leaving the cluster. The typical time the first
particle needs to separate from the cluster is simply given by $T =
q_0^{-1}$. This time already determines the average gap between two 
particles in the low-density regime. Therefore the average density,
$\rho_{free}$ in the free-flow regime is given by
$\rho_{free} = 1 /(T+1) =  q_0/(1+q_0)$. The number of moving
particles and therefore the flux follows from particle conservation
(see also \cite{Barlo}).

Both solutions for $J(\rho)$ 
coexist for $\rho_{free} \leq \rho \leq 1/2$. For finite
systems it is possible that the large cluster dissolves due to
fluctuations. In free flow states, fluctuations are
absent and therefore no spontaneous cluster formation takes place at
densities $\rho \leq 0.5$.
However, in the thermodynamic limit, any random initial configuration
leads to a jammed stationnary state.

After this short discussion of the system with PBC,
we proceed with the open chain. OBC
are implemented as follows : at the left boundary
particles with velocity one are introduced with probability $\alpha$ if the 
first site of the chain is empty. At the right end the particles
leave the chain from site $L$ with probability $\beta$, irrespective
of their velocity. 

The values of $\alpha$ and $\beta$ determine for given values of $L$
and $q_0$ the bulk behavior of the system. First we notice that for
$\beta = 1$ the only stochastic element is due to the left boundary,
i.e. we recover the deterministic asymmetric exclusion
process. Therefore the exact result for the flow
\cite{EvansdeGier} is given by
$J(\alpha) = \alpha /(1+\alpha)$.
This result for the flow is expected to be valid not only for $\beta
=1$ but in the whole low density phase, because for low bulk densities
the flow is controlled by the input of particles. 

Next, we consider the case of large bulk densities, i.e. we study the
system's performance for large values of $\alpha$ and $\beta \ll 1$. For
this case we expect to find frequently a compact cluster of particles
at the right boundary. Therefore the mean time interval for a particle
leaving the 
system is the sum of the average waiting time $T_L = 1/\beta$ of the
last particle at site $L$ and the average time which passes until the
last site is occupied again, $T_J = 1/q_0$. We find immediately
that for large $\alpha \approx 1$ and $\beta \ll 1$ the flow is
given by
   
\begin{equation}
J(\beta) = \frac{1}{T_L+T_J} = \frac{q_0 \beta}{q_0 + \beta}.
\label{eq_jb}
\end{equation}

\begin{figure}[h]
\centerline{\psfig{figure=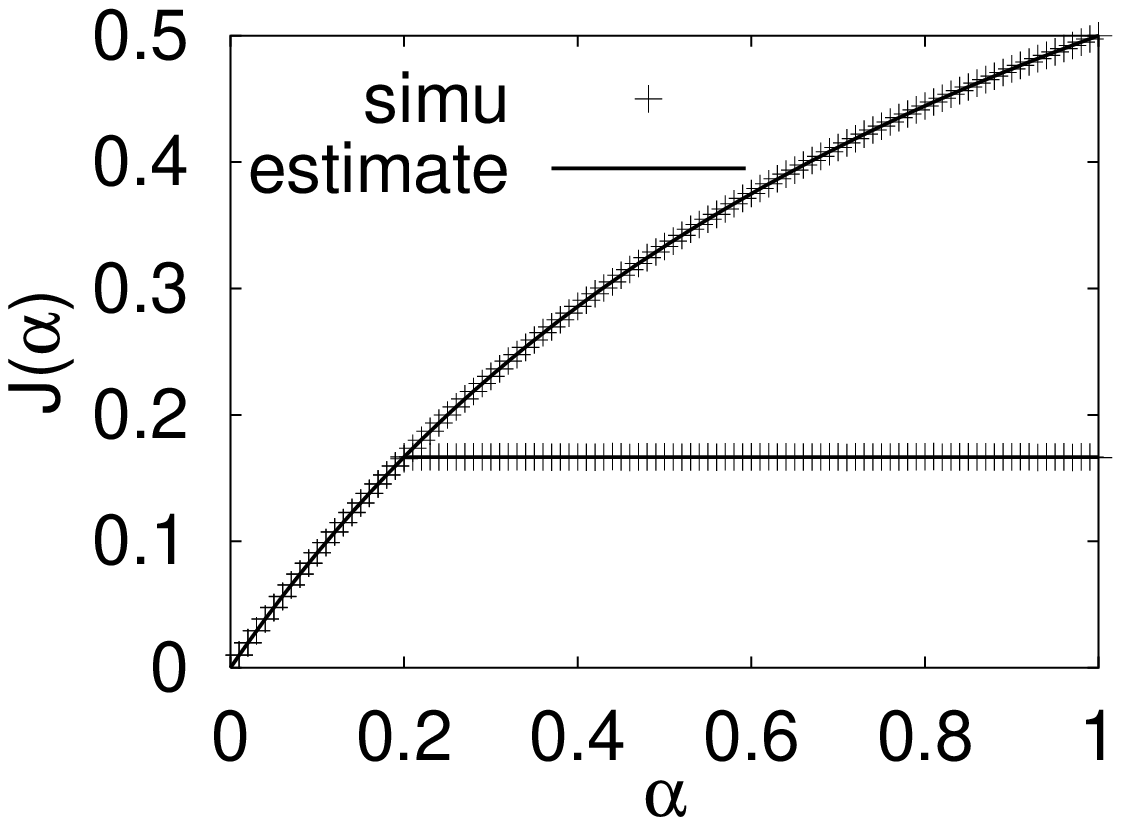,height=5cm}}
\centerline{\psfig{figure=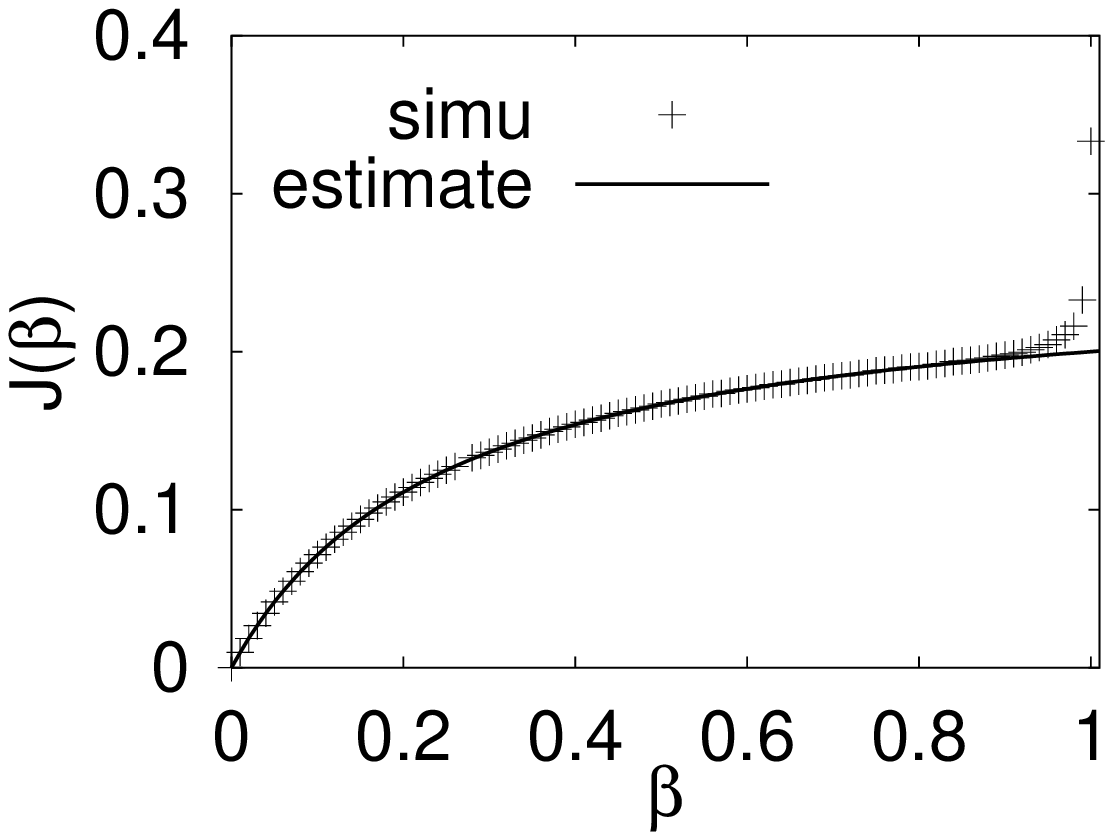,height=5cm}}
\caption{\protect{Comparison of the estimated flow with simulation 
results. The simulation results have been obtained for a chain of
length $L=500$ and $q_0=0.25$. {\bf (a)} The Figure shows the 
perfect agreement of simulation results and the estimate 
$J(\alpha)$. We used $\beta = 1.$ and $\beta = 0.5$ as output probabilities.
{\bf (b)} } $\beta$ dependence of the flux ($\alpha = 0.5$). The
estimated flux and the simulation results agree for sufficiently small
values of $\beta$. Simulations for different lengths of the chain
indicate that the deviations for  $\beta$ near one are due to finite size 
effects.}
\label{fig_openflow}
\end{figure}

This simple scenario is only valid if the density of holes fed into
the system is very low. For larger values of $\beta$ we expect that it is also
possible that the cluster separates from the right boundary. In this
case the density close to the exit is determined by the outflow from a
cluster. The temporal headway $t$ between two consecutive 
particles is distributed via $P_h(t) = q_0 (1-q_0)^{t-1}$ where  $t \geq
1$. Now we calculate the typical amount of time which is needed from
the arrival of a particle (at $t=0$) to the arrival of the second particle at 
($t=T$) at site $L$. Here we have to distinguish two cases: $(i)$ the
particle arrives at site $L-1$ {\em before} the first particle has left the
system or $(ii)$ that it arrives later, such that it reaches the last site
without being blocked by the particle ahead. The probability that a
particle gets blocked if it arrives at $t$ is given by $P_b (t) =
(1-\beta)^t$ and that it can pass without blockage by 
$P_f (t) = \sum_{\tau=1}^{t} \beta (1-\beta)^{t-1}$.

This leads to the average value of $T$  given by 
\begin{eqnarray}
T &=& \sum_{t=1}^{\infty} P_h(t) \left[ P_f (t) (t+1) +  P_b(t) (
t + \frac{1}{\beta} + \frac{1}{q_0}) \right] \nonumber \\
  &= &\frac{1}{\beta} + \frac{1}{q_0} = J(\beta)^{-1}.
\end{eqnarray}
Also for this scenario we recover the same value
(\ref{eq_jb}) of the 
flow. These two kind of configurations are generic if the output
probability controls the capacity of the system, i.e. in the presence
of clusters. Therefore we expect to find a unique functional dependence of
the flow, if the system is controlled by the capacity of the exit.
The transition between the $\alpha$ or $\beta$ controlled parameter 
regime is expected to be located at 
\begin{equation}
\beta_t (\alpha) = \frac{ q_0 \alpha}{(1+\alpha) q_0-\alpha}
\end{equation}
where both fluxes agree.

\begin{figure}[h]
\centerline{\psfig{figure=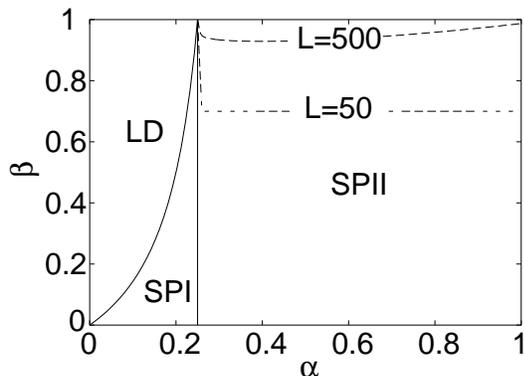,height=5cm}}
\caption{\protect{Phase diagram of the open system for $q_0 = 0.25$.
The phases are in order: low density phase (LD), striped  phase I
(SPI), and striped phase II (SPII). 
At the line $\alpha = q_0$ one observes a
constant density profile. The dashed lines show parameter regimes
where in finite systems a systematically larger flux has been
observed. }}
\label{fig_phase}
\end{figure}

Comparison of our prediction for the flow with simulation results
shows excellent agreement (see fig.~\ref{fig_openflow}).
Simulations for different lengths of the
chain show that the differences between our predictions and the 
simulations systematically vanish for larger system sizes.
The origin of  these finite size effects, i.e. the enhancement of
the flow for large $\beta$,
is the possibility to find occasionally cluster-free
micro-states. We indicated this by the broken lines in the phase diagram
(Fig. \ref{fig_phase}).
The absence of maximal current phase in the thermodynamic limit
is related to the non-analyticity of the fundamental
diagram \cite{Kolo98}.

\begin{figure}[h]
\centerline{\psfig{figure=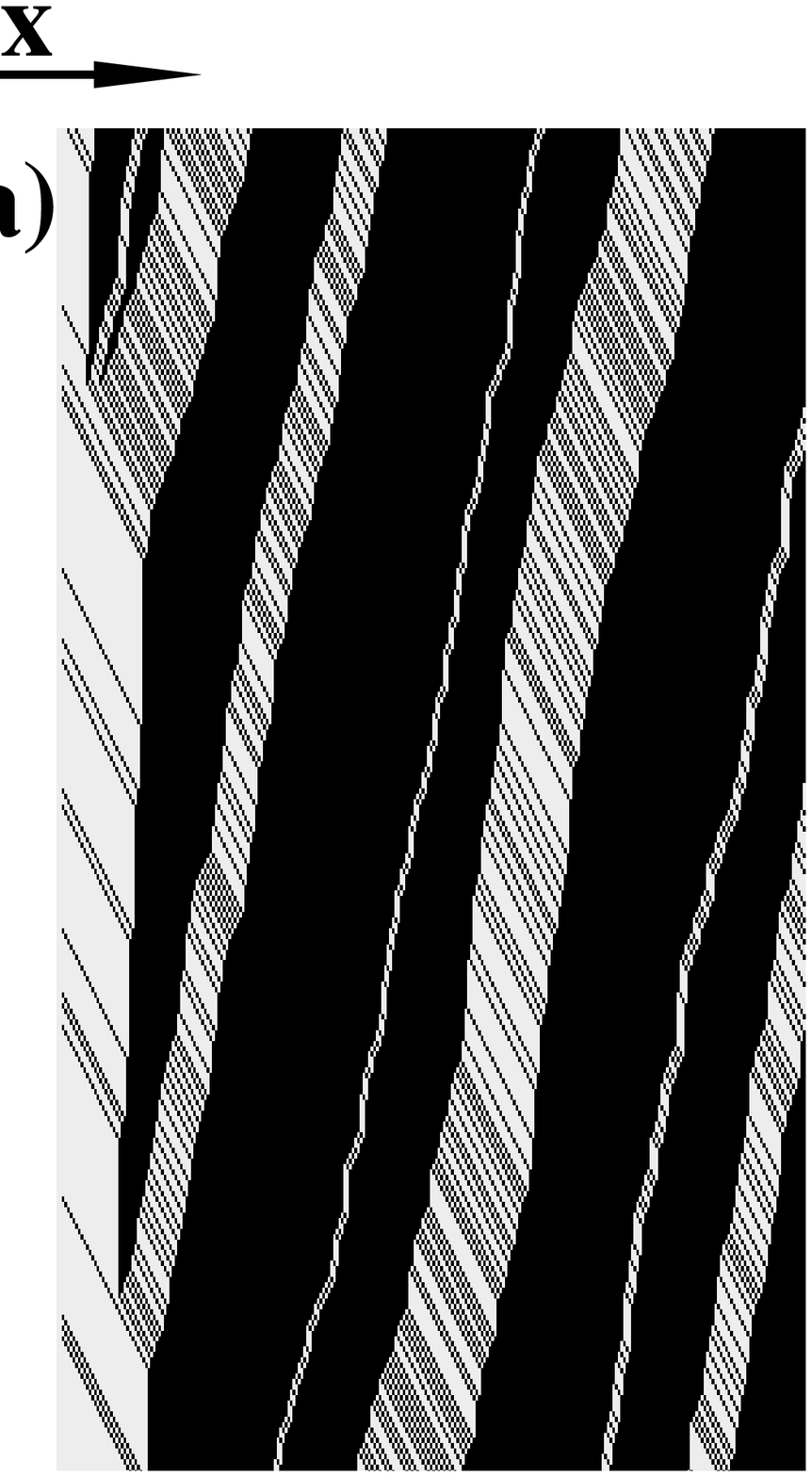,height=6.5cm}
\psfig{figure=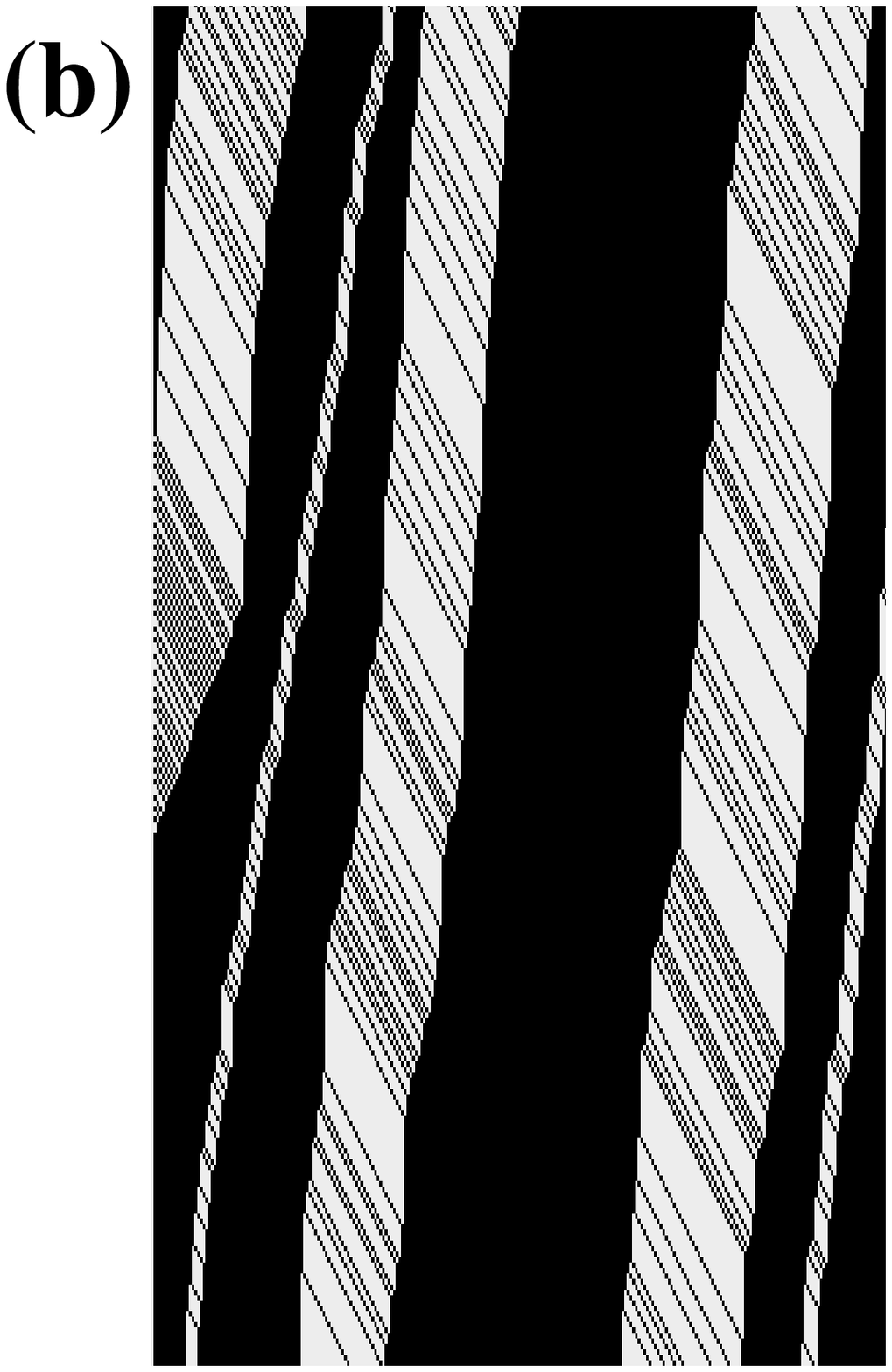,height=6.5cm}}
\caption{\protect{Typical space-time plots in the striped phases. The
figure shows the first $\sim 400$ sites of a chain of length $L =
1,000$.  {\bf(a)} For $SPI$ the size of the last clusters decreases
rapidly close to the entrance. ($\alpha = 0.1, \beta = 0.1, q_0 = 0.4$) 
{\bf(b)} For $SPII$ the efficient particle reservoir at the
entrance of the system augments the size of the leftmost cluster.}}
\label{fig_st}
\end{figure}
We proceed characterizing the typical microscopic structure of the 
stationary state, in different parameter regimes. We distinguish three
different domains, i.e. a
{\em low density} domain (LD) for $\beta >\beta_t
(\alpha)$ (LD), and two high density domains for $\beta <\beta_t
(\alpha)$. Both high density domains  correspond to a {\em striped
phase} (SP), but they  exhibit different behaviors in the vicinity of
the entrance, depending whether $\alpha < q_0$ (SPI) or $\alpha \geq q_0$
(SPII). In the LD phase we observe the typical two domain
structure known from the asymmetric exclusion process. Clusters which
eventually separate from 
the high density domain at the right dissolve in a few time steps. This
dissolution leads to the localization of the high density domain at
the right. The system takes a bulk density $\rho_b(\alpha) =
J(\alpha)$. At site $L$ the density is given $\rho_L =
J(\alpha)/\beta$. Fits of several density profiles follow the 
pure exponential form $\rho(n) = \rho_b(\alpha) 
+ [\rho_L-\rho_b(\alpha)]\exp(-(L-n)/\xi)$.

\begin{figure}[h]
\centerline{\psfig{figure=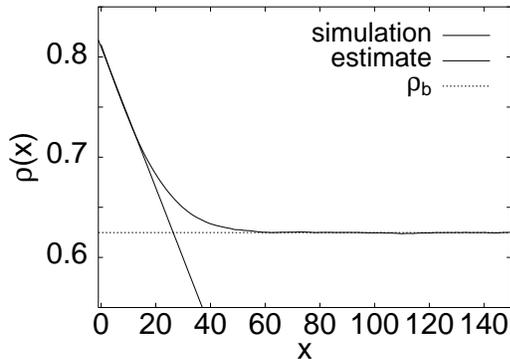,height=5cm}}
\caption{\protect{Density profile in the SPII region.
The numerical result (solid line) is compared with the theoretical
prediction (dashed line) based on an assumption of monodispersity
and parameterized by $t_f$ which has to be determined numerically.
}}
\label{fig_dp}
\end{figure}

In the high density region (SPI + SPII),
clusters are able to reach entrance.
This region can be described by considering that at the right
boundary, some free-flow segments are injected within a compact cluster.
Typically, the free-flow segments reduce to isolated holes for
$\beta<q_0$, whereas for $\beta>q_0$, they contain both empty sites
and particles with velocity one.
Both the free-flow segments and the compact clusters separating them
move backwards, producing a striped structure in spatio-temporal
plots (see fig.~\ref{fig_st}).
As long as it is far enough from the left boundary, a cluster has the same
in- and output rates, thus its width follows a non-biased random walk,
whose average is a constant.
However, there is a non-vanishing probability that the width of the
cluster crosses the zero value, leading to the coalescence of
the two neighboring free-flow segments, which will never be able
to separate again afterwards. The excess particles are distributed
among the neighboring clusters. Thus, as the stripes 
move towards the entrance, their widths increase.
However, the average density remains a constant equal to the density
selected on the left boundary $\rho = J(\beta)/\beta$.
When the clusters arrive near the left boundary and are exposed
to the entrance flux, their width suddenly obeys a {\em biased}
random walk. If $\alpha > q_0$ (phase SPII),
the width increases on average
and the high input rate stabilizes the cluster until it reaches the boundary
(fig.~\ref{fig_st}-b).
If clusters and free-flow segments at a given position were monodisperse, 
then the sudden rate change would induce a linear density
profile on the left, with a crossover to the constant bulk value
occurring at site
$\tilde{n} = \alpha t_f /(1+\alpha)$
where $t_f$ is the average time interval during which free-flow is
observed in the first site, i.e. the typical time between the
disappearance of a jam and the arrival of the next. Due to the
coarsening of the striped pattern in upstream direction one observes
a subextensive growth of $t_f$ with increasing system size.  
The density in the first site is determined by noticing that
the flux (\ref{eq_jb}) is also equal to $j = \alpha(1-\rho_1)$.
If we measure $t_f$ numerically,
we find a quite good agreement between our estimate and  the measured
density profile 
(see fig.~\ref{fig_dp}), except that the crossover region is in fact
very large due to size dispersion of clusters and free-flow segments.

If $\alpha < q_0$ (phase SPI), the width of the leftmost cluster
decreases on average
due to the reduced capacity of the entrance.
Some clusters will even not be able to reach the entrance,
as seen on fig.~\ref{fig_st}.
As long as we are far from the transition line $\beta_t(\alpha)$,
still, clusters are on average large enough so that most of them
reach the entrance and the depletion region remains localized
on the left.
The calculation done in SPII for the density profile is  still valid.
The only difference is that now the linear part has a positive slope.
However, as we approach the transition line $\beta_t(\alpha)$,
more and more clusters do not reach the boundary.
When a given cluster disappears before reaching the left end,
the next one is exposed earlier to the low incoming flux,
and thus the depletion region may invade the whole system.
The line separating the depletion region from the constant
density region becomes delocalized over the whole system
as the transition line is approached.

To conclude, we analyzed a DLG exhibiting meta-stable
states. The structure of the phase diagram was explored by simulations
as well as by a phenomenological description. The analytical
predictions for each phase agree extremely well with the simulation results.
The most characteristic feature is the {\em spontaneous}
formation of simultaneously existing clusters for restricted outflow.
The coarsening of these clusters as they flow backwards is due
to the stochasticity of the model.
We stress that this striped pattern survives in the thermodynamic
limit. We expect that this behavior and more generally the structure
of the phase diagram are generic for DLG with 
branched fundamental diagrams
- this branched structure being a signature of meta-stability.
Moreover we established, in the limit of large system size,
the absence of a maximal current phase.

Besides a phenomenological similarity with the clogging
phenomena in granular flow \cite{granular},
the practical relevance can be illustrated by the example of traffic
flow, where the importance of boundary induced phase transitions
is well accepted \cite{lee_treiner}.
 Although we chose a rather simplistic model and do not aim
to give a complete description of traffic flow, our results
can be related to experimental studies that revealed
the existence of parallel
existing clusters on highways \cite{Kerner961}.
These experimental results now can be 
directly related to the existence of meta-stable states.
Moreover, our result shows that jams can be observed far upstream
even when boundary induced.
Another
interesting feature are the important finite size effects, which
probably can be used for a systematic flow optimization 
\cite{optimal}.\\

{\bf Acknowledgments}: We thank Robert Barlovic, Andreas
Schadschneider, and J.~Krug for useful discussions. L.~S. acknowledges
support from the Deutsche Forschungsgemeinschaft under Grant No. SA864/1-1.

\bibliographystyle{unsrt}

\end{document}